\documentclass[3p,sort&compress]{elsarticle}

\usepackage{lineno,hyperref}
\usepackage{amssymb}
\usepackage{amsmath}

\modulolinenumbers[5]

\journal{Physics Letters A}

\begin{document}

\begin{frontmatter}

\title{{\em A Priori} Which-Way Information in \\ Quantum Interference with Unstable Particles}

\author[Wabash,Purdue]{D.E.~Krause\corref{cor1}}  \ead{kraused@wabash.edu}

\address[Wabash]{Physics Department, Wabash College, Crawfordsville, IN 47933, USA}
\address[Purdue]{Department of Physics and Astronomy, Purdue University, West Lafayette, IN 47907, USA}

\author[Purdue]{E.~Fischbach}

\author[Avon]{Z.J.~Rohrbach}
\address[Avon]{Avon High School, 7575 East 150 South,
Avon, IN 46123, USA}

\cortext[cor1]{Corresponding author}

\begin{abstract}
If  an unstable particle  used in a two-path interference experiment decays before reaching a detector, which-way information becomes available which reduces the detected interference fringe visibility ${\cal V}$.  Here we argue that even when an unstable particle does {\em not} decay while in the interferometer,  {\em a priori} which-way information is still available in the form of path predictability ${\cal P}$ which depends on the particle's decay rate $\Gamma$.  We further demonstrate that in a matter-wave Mach-Zehnder interferometer using an excited atom  with an appropriately tuned cavity,  ${\cal P}$ is related to ${\cal V}$ through the duality relation ${\cal P}^{2} + {\cal V}^{2} = 1$.

\end{abstract}


\end{frontmatter}

%\linenumbers

\section{Introduction}

The interference of particles is a signature quantum mechanical effect that has been demonstrated with photons \cite{Grynberg}, neutrons \cite{Werner book}, atoms,  and molecules \cite{Cronin}.  While the interference of classical waves has a straightforward interpretation, the interference of particles appears more mysterious since the effect arises from the summation of probability amplitudes rather than physical field amplitudes. Since the early days of quantum mechanics and the famous debates between Bohr and Einstein, it came to be realized that the amount of quantum interference one observes depends on the quantity of information available on the path the particle takes \cite{Pais,Greenstein}.  If one knows which way (i.e., which path) the particle takes, there is no interference, while if there is no which-way information, one observes complete interference.   The effect of  having partial which-way information on quantum interference  has been investigated for the double-slit experiment \cite{Wootters Zurek}, and   for  Mach-Zehnder photon \cite{Glauber} and neutron  \cite{Greenberger Yasin} interferometers.  

 Let us now quantify this discussion. A common measure of the amount of interference is  the interference fringe visibility,  ${\cal V}$, which is given by 
\begin{equation}
{\cal V} = \frac{I_{\rm max} - I_{\rm min}}{I_{\rm max} + I_{\rm min}},
\label{V}
\end{equation}
where $I_{\rm max}$ and $I_{\rm min}$ are the adjacent maximum and minimum observed intensities as some variable (e.g., a phase shift in one path) is varied.   The visibility measures the amount of contrast in the interference, with fully coherent beams leading to  maximum contrast (${\cal V} = 1$), while loss of coherence reduces the fringe contrast.  

 To quantify  which-way information, a number of different measures have been used in the literature.  Here we will follow Englert \cite{Englert} and distinguish between two types of which-way information. If the interfering particle becomes entangled with another quantum system such that the path taken is ``marked''  in some way \cite{Cronin}, one can use the path distinguishability ${\cal D}$ as a measure of the which-way information \cite{Jaeger,Englert,Schwindt,Durr AJP}.  
 For example, a which-way detector may interact with the particle or the interfering particle could emit radiation (e.g.,  thermal radiation from a $^{60}$C molecule or a photon from an excited atom decaying to the ground state) while in the interferometer, giving away its position.  In  these situations, the interfering particle becomes entangled with another  quantum system which leads to coherence loss of the interfering beams, and hence a loss of interference contrast. The distinguishability ${\cal D}$ can be expressed in terms of the reduced density operator of the detector or other entangled system \cite{Englert}, but its exact form is not needed here.
 It has been shown that for  two-path interference, ${\cal D}$ and ${\cal V}$ satisfy the duality relation \cite{Jaeger,Englert,Schwindt,Durr AJP}
 \begin{equation}
 {\cal D}^{2} + {\cal V}^{2} \leq 1,
 \label{D V}
 \end{equation}
 where the equality holds if the interfering particle and marking system are described by pure states.
 
 If, on the other hand, information is available which allows one to predict which  path a particle will take before entering the interferometer, (e.g., from an asymmetry in the two paths),  which-way information may be quantified using the path predictability  ${\cal P}$ given by \cite{Glauber,Greenberger Yasin,Jaeger,Englert,Schwindt,Durr AJP},  
\begin{equation}
{\cal P} = |P_{1} - P_{2}|,
\end{equation}
where $P_{i}$  is the probability that a particle  which reaches the detector took  path $i$.  One then
finds that ${\cal P}$ and ${\cal V}$ satisfy \cite{Glauber,Greenberger Yasin,Jaeger,Englert,Schwindt,Durr AJP}
\begin{equation}
{\cal P}^{2} + {\cal V}^{2} \leq 1.
\label{P V}
\end{equation}
In contrast to ${\cal D}$, which represents {\em a posteriori} which-way information (i.e., information gained {\em after} the interfering particle has been sent into interferometer), ${\cal P}$ represents {\em a priori} which-way information, available {\em before} the particle is sent into the interferometer \cite{Jacques}.  Clearly one can combine the two types of which-way information to yield a single inequality, but this is not needed for our purposes.

Whatever the case, Eqs.~(\ref{D V}) and (\ref{P V}) tell us that any type of which-way information will reduce the interference contrast.   
The equality holds for coherent beams, in which case knowing which way the particle traveled (${\cal D} = 1$ or ${\cal P} = 1$) causes the interference to disappear (${\cal V} = 0$), while having no which-way information (${\cal D} = 0$ or ${\cal P} = 0$) yields the maximum interference contrast (${\cal V} = 1$).  Eq.~(\ref{D V}) has been confirmed by experiments using photons \cite{Jacques} and atoms \cite{Durr}, while Eq.~({\ref{P V}) has been verified with  photons \cite{Liu}, and neutrons \cite{Zawisky}.  In a related set of experiments conducted by Summhammer, et al. \cite{Summhammer},  absorbing foils, choppers, and an absorbing lattice were placed into one of the paths of a neutron interferometer to investigate how the interference visibility was affected by stochastic (completely unpredictable) or deterministic (completely knowable) absorption of neutrons from one of the beams.  Finally, the impact that  internal particle dynamics can have on interference contrast has been investigated recently by Zych, et al., \cite{Zych} and Banaszek, et al. \cite{Banaszek}.

In nearly all previous work, the interfering particles were considered to be stable.  [Although neutrons are unstable, their lifetimes  are so long ($\tau_{\rm neutron} \simeq 882$~s) that they were considered to be stable in the analysis of neutron interferometer experiments.]   Bramon, et al. \cite{Bramon} and Hiesmayer and Huber \cite{Hiesmayr} have examined a duality relation analogous to Eq.~(\ref{P V}) in the neutral kaon system, which is unstable.  Several authors have studied the impact on an interference pattern when an excited atom decays en route to the detector theoretically~\cite{Sleator,Facchi,Takagi} and experimentally \cite{Pfau}, which  probes the  {\em a posteriori} duality relation~(\ref{D V}) assuming a symmetric setup.  However,  they did not consider in detail the case where the unstable particle does not decay  while in the interferometer and no which-way information is gained during the experiment. The purpose of this paper is to fill in this gap by  investigating the  interference of  {\em undecayed} unstable particles.  (Of course, an unstable particle has, by definition, not decayed, but we use this redundant terminology to distinguish  our situation from the case of interference of an unstable particle's decay products.)  We show that the path predictability ${\cal P}$ describing unstable particles for a symmetric 2-path interference setup is not generally zero as for stable particles, but depends on the particle decay rate $\Gamma$, and that it  satisfies the {\em a priori} duality relation~(\ref{P V}).   

One can envision  three situations where non-trivial path predictability arises with unstable particles, all of which would yield no which-way information if stable particles were used.   In a double slit experiment,  an unstable particle travels with the same speed (momentum)  over two paths of different lengths, so the particle traveling the shorter path will have a greater chance of reaching the detector compared to the  particle taking the longer path. (Note we are using classical trajectory language as an intuitive guide which should not be taken literally.)  In a Colella, Overhauser, and Werner (COW)-type experiment \cite{COW}, the unstable particle  travels with different speeds (momenta) over two paths of equal length, so the particle taking the quicker path will have a greater a chance of reaching the detector compared to the slower path.    (Using the COW experiment with unstable particles to investigate the gravitational equivalence principle has been discussed recently by Bonder, et al. \cite{Bonder}.)
However, one can show that systematic effects (e.g., finite coherence length) in double slit or COW experiments will affect the visibility of the interference  in the same manner as the instability and so will make observations of the instability effect difficult \cite{Krause unpublished}.  Instead, here we will focus  on a third situation which does not suffer from these difficulties,  a Mach-Zehnder-like atom interferometer with excited atoms.  By placing an appropriately tuned cavity in one of its arms to modify the decay rate, the probability of the atom reaching the detector via one path will change relative to the other.    We will then show that path predictability and interference visibility satisfy the duality relation~(\ref{P V}).

\section{Unstable Atom Mach-Zehnder Interference with a Cavity}

\subsection{Setup}

In a typical matter wave Mach-Zehnder interferometer (Fig.~\ref{cavity setup figure}), 
\begin{figure}[tbp]
\begin{center}
\includegraphics[width=90mm]{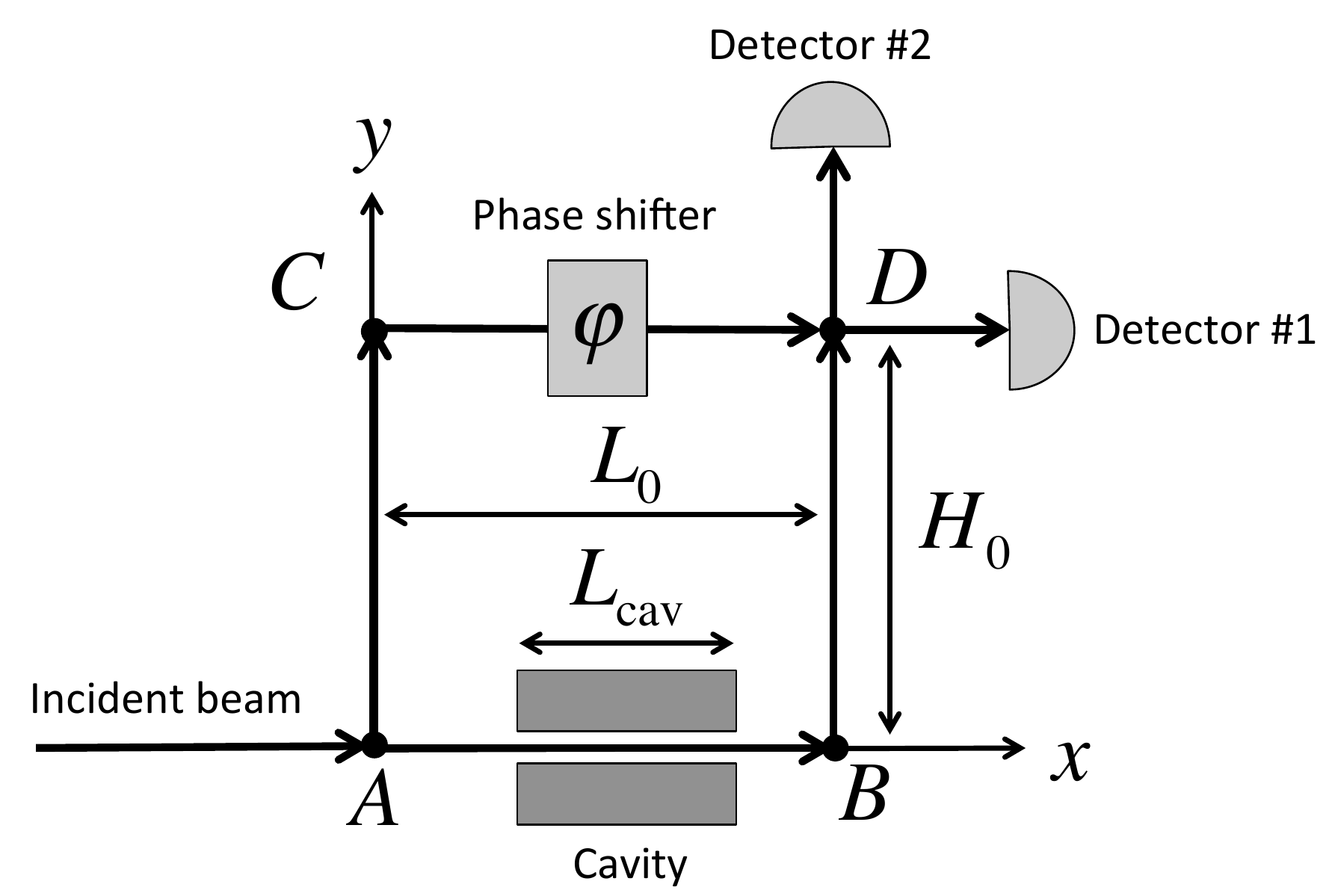}
\caption{A schematic of a Mach-Zehnder-type atom interference experiment using unstable atoms with a cavity of length $L_{\rm cav}$ placed in the arm $AB$.  The interferometer lies in the horizontal plane, and a phase shifter is present in the path $CD$ to modulate the interference.}
\label{cavity setup figure}
\end{center}
\end{figure}
a beamsplitter at $A$ divides an incident beam of particles into two beams which are then reflected, via two mirrors $B$ and $C$, into a second beamsplitter $D$ which then can send particles to either detector \#1 or \#2.  In this setup, the paths $ABD$ and $ACD$ are of equal length, and we assume no significant external potentials, so the speeds of the particles are the same and they will take the same time to traverse each path from $A$ to $D$.  A phase shifter is  inserted into  the arm $CD$ to create a variable phase difference between the two paths which modulates the probability that the particle will reach detectors \#1 and \#2.  Since the two paths are completely symmetric for both stable and unstable particles, we expect no difference in the observed interference with the conventional Mach-Zehnder configuration.

To observe a non-trivial modification of the detection probabilities due to particle instability for this setup, we need to modify the decay rate in one of the beams.  Fortunately, this is possible for atoms since  the decay rate of an atomic excited state is not an intrinsic property of  the atom.  Rather, the decay process arises from the atom's interaction with the quantum electromagnetic field, and so the decay rate can be manipulated by modifying the field's vacuum state by placing the atom in a cavity \cite{Haroche PT}.  A number of experiments have demonstrated the enhancement and suppression of an excited atom's decay rate in appropriately tuned cavities \cite{CQED experiments,Jhe}.   To take advantage of this effect, let us insert  a cavity of length $L_{\rm cav}$  into the the arm $AB$ as shown in Fig.~\ref{cavity setup figure}.  While inside the cavity, the vacuum decay rate of an excited atom $\Gamma$ changes to the cavity-modified rate $\Gamma_{\rm cav}$, which means that the probability that the excited atom will arrive at $D$ undecayed via path $ABD$ will differ from the atom taking the path $ACD$.  We will now show that the visibility ${\cal V}$ observed at the detectors will depend on the ratio $\Gamma_{\rm cav}/\Gamma$, and will be related to the path predictability ${\cal P}$ through the duality relation~(\ref{P V}).

\subsection{Propagation Probability Amplitude}

To calculate the detection probability for excited particles traversing the Mach-Zehnder interferometer, we first need to find the propagation probability amplitude for unstable particles.  We purposefully use a simple phenomenological model to describe unstable particles in order to encompass the widest range of possibilities while also avoiding inessential details of the decay process that would tend to obscure the fundamental issues we wish to address.   The  non-relativistic Schr\"{o}dinger equation describing the interaction-free center of mass motion of an atom in an excited state will be written as
\begin{equation}
i\hbar\frac{\partial\Psi(\mathbf{r},t)}{\partial t} = \left(mc^{2} - i\frac{\hbar\Gamma}{2} - \frac{\hbar^{2}}{2m}\nabla^{2}\right)\Psi(\mathbf{r},t),
\label{S E}
\end{equation}
 $m$ is the mass of the atom and $\Gamma$ is the decay rate of the excited state. We will consider only stationary beam experiments, so we will follow Greenberger and Overhauser \cite{G O} and search for  solutions of Eq.~(\ref{S E}) of the form 
\begin{equation}
\Psi_{E}(\mathbf{r},t) = \psi_{E}(\mathbf{r})e^{-iEt/\hbar},
\label{Psi}
\end{equation}
 where $E$ is a real energy.  Substituting Eq.~(\ref{Psi}) into Eq.~(\ref{S E}) gives the time-independent 
 Schr\"{o}dinger equation, which can be rewritten as
\begin{equation}
\nabla^{2}\psi_{E}(\mathbf{r}) = -\tilde{\mathbf{k}}^{2}\psi_{E}(\mathbf{r}),
\label{k psi0}
\end{equation}
where the magnitude of the complex wave vector $\tilde{\mathbf{k}}  = \tilde{k}\hat{\mathbf{k}}$ is
\begin{equation}
|\tilde{\mathbf{k}}|^{2} = \tilde{k}^{2} \equiv (k + i \kappa)^{2}   \simeq \frac{p^{2}}{\hbar^{2}} + i\frac{m\Gamma}{\hbar}.
\label{k}
\end{equation}
(Note that $\hat{\mathbf{k}}$ is the real unit vector in the direction of  propagation.) Here $k$ and $\kappa$ are the real and imaginary parts of $\tilde{k}$, $p$ is the particle's momentum, and the non-relativistic limit of the kinetic energy, $E - mc^{2} \simeq p^{2}/2m$, was used.  Since $E$ is assumed to be real,  Eq.~(\ref{k}) can be used to show that
\begin{eqnarray}
k   & \simeq &    {p/\hbar} = 1/\lambda, \\
 \kappa & \simeq &  m\Gamma/2p = 1/2\ell,
 \end{eqnarray}
where $\lambda$ is the de~Broglie wavelength of the particle and $\ell = p/m\Gamma$ is the average distance the particle with momentum $p$ travels before decaying.  Here  we assumed the solutions  have a well-defined wavelength (i.e., momentum) so $\ell \gg \lambda$, which, when combined with the non-relativistic limit, implies the hierarchy of energy scales $mc^{2} \gg p^{2}/2m \gg \hbar\Gamma$.

The general solution to the unperturbed time-independent wave equation, Eq.~(\ref{k psi0}), can then be written as
\begin{equation}
\psi_{E}(\mathbf{r})  = Ae^{i\tilde{\mathbf{k}}\cdot\mathbf{r}} = 
A\exp\left[i\frac{\mathbf{p}\cdot\mathbf{r}}{\hbar}\left(1 + i\frac{m\hbar\Gamma}{2p^{2}}\right)\right].
\label{E0 psi}
\end{equation}
where $A$ is a constant.  This implies that the probability amplitude that the unstable particle will travel from $\mathbf{r}_{0}$ to $\mathbf{r}_{f}$ can be written as
\begin{equation}
\psi_{0 \rightarrow f} = Ae^{ips/\hbar}e^{-s/2\ell},
\label{free psi}
\end{equation}
where $s$ is the total distance traveled  along its path $\mathbf{r}_{0} \rightarrow \mathbf{r}_{f}$.  This is just what one would expect intuitively: the factor $e^{ips/\hbar}$ is the usual   amplitude for freely propagating  stable particles, and the term $e^{-s/2\ell}$ accounts for the depletion of the beam arising from particle decays.  

While Eq.~(\ref{free psi}) describes the propagation amplitude for a freely propagating unstable particle, it is straightforward to generalize these results for particles traveling in a potential $V(\mathbf{r})$ using the approach of Ref.~\cite{G O}.  One finds that
\begin{equation}
\psi_{0 \rightarrow f}  \simeq Ae^{ips/\hbar}e^{-s/2\ell}e^{i\tilde{\phi}_{0f}},
\end{equation}
where
\begin{equation}
\tilde{\phi}_{0f} \equiv -\frac{m}{\hbar p}\left[1- i\left(\frac{\lambda}{2\ell}\right)\right]\int^{\mathbf{r}_{f}}_{\mathbf{r}_{0}}V(\mathbf{r})\,ds,
\label{complex phase}
\end{equation}
and the integral is over the path taken by the particle when $V(\mathbf{r}) = 0$ \cite{Krause unpublished}.

\subsection{Detection Probability}

To calculate the probability that an excited atom arrives at detector \#1 undecayed,  we will assume for convenience that the transmission and reflection amplitudes for the beamsplitter are ${\cal T}_{\rm BS} =  1/\sqrt{2}$ and ${\cal R}_{\rm BS} =i/\sqrt{2}$, respectively, while the reflection amplitude for the mirrors is ${\cal R}_{\rm M} =-1.$  Using these amplitudes together with Eq.~(\ref{free psi}), we find  the probability amplitude that the unstable particle with momentum $p$ arrives at detector \#1 via the path $ABC$ is  given by
\begin{equation}
\psi_{1, ACD} = -\frac{i}{2}e^{ip(H_{0}+ L_{0})/\hbar}e^{i\varphi}e^{-(H_{0}+ L_{0})/2\ell},
\label{psi ACD}
\end{equation} 
where $H_{0}$ and $L_{0}$ are the lengths of the arms of the interferometer shown in Fig.~\ref{cavity setup figure}.  The  factor $e^{i\varphi}$ accounts for the action of the phase shifter.  Similarly, the probability amplitude that the particle reaches detector \#1 through the path $ABD$ is 
\begin{eqnarray}
\psi_{1, ABD} & = & -\frac{i}{2}e^{ip(H_{0}+ L_{0})/\hbar}e^{-[H_{0}+ (L_{0} - L_{\rm cav})]/2\ell}e^{-L_{\rm cav}/2\ell_{\rm cav}},
\label{psi ABD}
\end{eqnarray} 
where $\ell_{\rm cav} = p_{0}/m\Gamma_{\rm cav}$ is the mean distance the particle will travel in the cavity before decaying with the modified decay rate $\Gamma_{\rm cav}$.

Using Eqs.~(\ref{psi ACD}) and (\ref{psi ABD}), we find that the probability that the unstable particle reaches detector \#1 is given by
\begin{eqnarray}
P_{1}(\varphi) &=& \left|\psi_{1, ACD} + \psi_{1, ABD}\right|^{2},
 \nonumber \\
& = &   \frac{1}{4}e^{-(H_{0}+L_{0})/\ell}\left|e^{i\varphi}+ \exp\left[\frac{L_{\rm cav}}{2\ell}\left(1 - \frac{\Gamma_{\rm cav}}{\Gamma}\right)\right]\right|^{2}, 
\label{P1 cavity a}
\end{eqnarray}
where  $\ell/\ell_{\rm cav} = \Gamma_{\rm cav}/\Gamma$.  If we define
\begin{equation}
\theta_{\rm cav} \equiv \frac{L_{\rm cav}}{2\ell}\left(1 - \frac{\Gamma_{\rm cav}}{\Gamma}\right),
\end{equation}
 Eq.~(\ref{P1 cavity a}) can be rewritten into the form
 \begin{equation}
 P_{1}(\varphi) = \frac{1}{4}e^{-(H_{0} + L_{0})/\ell}\left(1 + e^{2\theta_{\rm cav}}\right)(1 + \mbox{sech}\, \theta_{\rm cav}\cos\varphi),
 \label{P1 cavity b}
 \end{equation}
which reduces to the correct result for stable particles, 
 \begin{equation}
 P_{1}^{\rm stable}(\varphi) = \frac{1}{2}(1 + \cos\varphi),
 \label{P1 stable}
 \end{equation}
 when the mean propagation length $\ell \rightarrow \infty$.  One can similarly calculate the probability that the unstable particle reaches detector \#2, obtaining
 \begin{equation}
 P_{2}(\varphi) = \frac{1}{4}e^{-(H_{0} + L_{0})/\ell}\left(1 + e^{2\theta_{\rm cav}}\right)(1 - \mbox{sech}\, \theta_{\rm cav}\cos\varphi).
 \label{P2 cavity}
 \end{equation}

\subsection{Visibility and Predictability}

Using the detection probability given by Eq.~(\ref{P1 cavity b}), the intensity observed by detector \#1 is
 \begin{equation}
 I_{1}(\varphi) = \frac{I_{0}}{1 + \mbox{sech}\, \theta_{\rm cav}}(1 + \mbox{sech}\, \theta_{\rm cav}\cos\varphi),
 \end{equation}
 where $I_{0} \equiv I_{1}(\varphi = 0)$.  The visibility of the interference pattern with the cavity is then
 \begin{equation}
 {\cal V} =  \mbox{sech}\, \theta_{\rm cav} =  \mbox{sech}\, \left[ \frac{L_{\rm cav}}{2\ell}\left(1 - \frac{\Gamma_{\rm cav}}{\Gamma}\right)\right],
 \label{V}
 \end{equation}
 which is unity when the cavity vanishes ($L_{\rm cav}  = 0$) or if  the cavity does not affect the decay rate ($\Gamma_{\rm cav} = \Gamma$).  
 
 It is now straightforward to show that these results satisfy the relation~(\ref{P V}).  The probability that the unstable particle which arrived at $D$ took a particular path  is determined by blocking one path and determining probability that the particle reaches the detector via the other path.  Taking the difference between the results gives the path predictability ${\cal P}$:
\begin{eqnarray}
{\cal P}
& = & \left|  \frac{| \psi_{1,ACD}|^{2} - |\psi_{1,ABD}|^{2}}{ |\psi_{1,ACD}|^{2} + |\psi_{1,ABD}|^{2}}\right|, \nonumber \\
&=& \tanh\left|\frac{L_{\rm cav}}{2\ell}\left(1 - \frac{\Gamma_{\rm cav}}{\Gamma}\right)\right|, \nonumber \\
& = & \tanh|\theta_{\rm cav}| 
\label{P}
\end{eqnarray}
 Notice the predictability uses {\em renormalized} probabilities to account for particles that decayed while in the interferometer; the predictability involves only those particles which actually reach the detector.
 If the decay is suppressed entirely while traveling in the cavity ($\Gamma_{\rm cav} = 0$),  ${\cal V} = \tanh(L_{\rm cav}/2\ell)$ and ${\cal P} = \mbox{sech}\,(L_{\rm cav}/2\ell)$.  If the cavity were tuned to enhance the decay such that $\Gamma_{\rm cav} \rightarrow \infty$, then ${\cal V} = 0$ and ${\cal P}= 1$ allowing one to predict with certainty that the atom took the path $ACD$ to reach the detector.

It is now clear from the identity $\mbox{sech}^{2}x + \tanh^{2} x = 1$ that Eqs.~(\ref{V}) and (\ref{P}) satisfy the duality relation~(\ref{P V}), and that  they have the same form as the unified description suggested by Bramon, et al. \cite{Bramon}. Using unstable particles in the Mach-Zehnder  experiment  with a cavity  gives  additional information on which path the unstable particle took to reach the detector compared to using stable particles.  The unstable particle is more likely to have taken the path which has the greater survival probability.

 Since the cavity  decay rate $\Gamma_{\rm cav}$ can be modified by changing the cavity (say, by mounting it asymmetrically to a rotating axis so that the atomic beam is in the cavity only periodically, or by sinusoidally varying the cavity's size), the interference fringe visibility can be modulated to ease detection.  Alternatively, one can apply a magnetic field to the excited atoms before they enter the interferometer to rotate the atoms as in the experiment by Jhe, et al. \cite{Jhe}.  The cavity decay rate depends on the orientation of the atom with respect to the cavity, so rotating the atoms in a regular manner would modulate the fringe visibility.  These two possibilities highlight the fact that a Mach-Zehnder-type experiment using excited atoms with a cavity provides more opportunities for manipulating the unstable which-way effect than the double slit or COW experiments using unstable particles.

\section{Conclusions}
 
In conclusion, we have extended the complementarity between {\em a priori} which-way information and interference fringe visibility, Eq.~(\ref{P V}), to interference with unstable particles.  We found that {\em a priori} which-way information results when using unstable particles  in a cavity Mach-Zehnder experiment compared with using stable particles. [Similar results can be found using   double slit and Colella-Overhauser-Werner (COW) experiments \cite{Krause unpublished}.] It is important to emphasize that we are considering the results of interference for {\em unstable particles that have not yet decayed.}  It may be surprising to see that the interference pattern of those unstable particles that survive the journey is different from the pattern produced by  their stable counterparts.  However, our results are consistent with the view of quantum mechanics as relating available information to observed phenomena, which lies at the heart of Eq.~(\ref{P V}).     It is quite possible  that using unstable particles in other interference problems will lead to similar new and interesting results.

 \section*{Acknowledgements}

We   thank Yuri Bonder, Hector Hern\'{a}ndez-Coronado, and Daniel Sudarsky for extensive discussions which provided  crucial insights that led to this work.  We also   thank Samuel Werner for illuminating conversations.



%% References

\end{document}